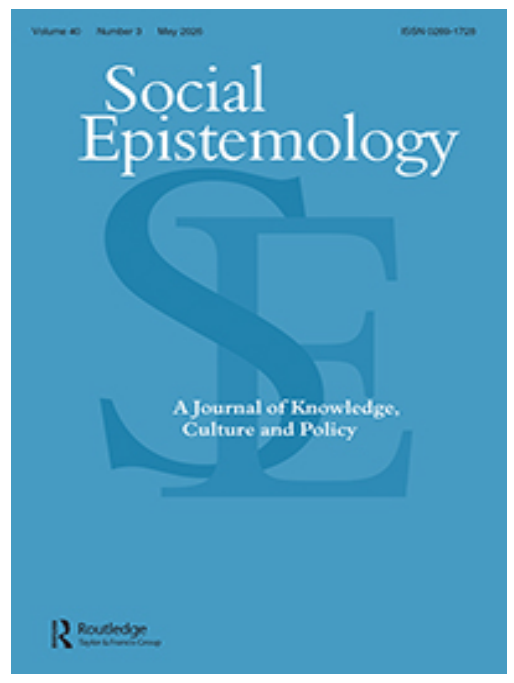

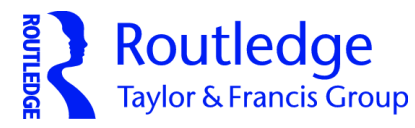

# Critically Engaged Pragmatism: Scientific Norm and Social, Pragmatist Epistemology for AI Science Evaluation Tools

**Carole J. Lee**

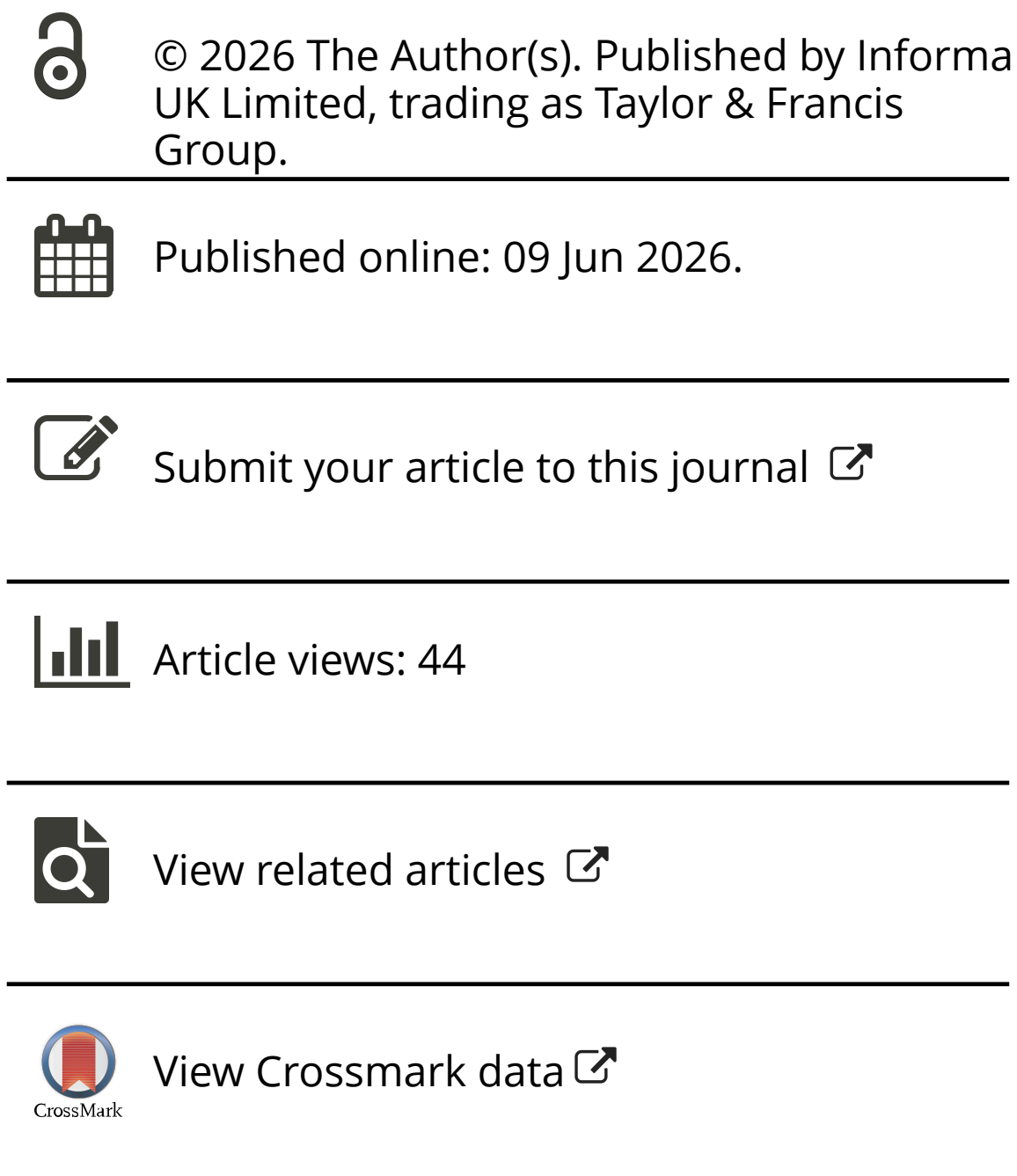

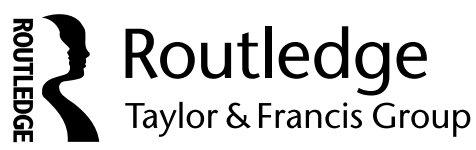

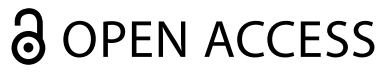

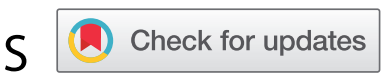

# Critically Engaged Pragmatism: Scientific Norm and Social, Pragmatist Epistemology for AI Science Evaluation Tools

Carole J. Lee

Department of Philosophy, Center for an Informed Public, University of Washington, Seattle, WA, USA

**ABSTRACT**

AI science evaluation tools aim to assess research credibility. As with traditional metrics such as impact factors, their edicts can be decontextualised and repurposed in problematic ways. To address this, I propose Critically-Engaged Pragmatism as a scientific norm enjoining scientific communities to scrutinise the purposes and purpose-specific reliability of AI science evaluation tools. To foster Critically Engaged Pragmatism, creators of AI science evaluation tools should transparently and fully report design, training, and benchmarking details to facilitate assessments of purpose-specific reliability, liability to different types of error, and bias. What count as best practices for the transparent reporting of AI science evaluation tools should be updated as new forms of error, bias, and gamesmanship are discovered. Under this framework, AI science evaluation tools are not objective arbiters of scientific credibility. Rather, they are the object of critical discursive practices that ultimately ground the credibility of scientific communities.

## 1. Introduction

Tools for automating research evaluation besiege the scientific community at a time of crisis for canonical mechanisms of credibility assessment.

The proliferation of papers is outpacing peer review capacity, putting enormous strain on and reconfiguring the normative structure of science (Lee 2022). Since 1990, the number of papers published in journals has grown exponentially (Dong et al. 2017), with preprints following a similar pattern of growth since arXiv's inception in 1991 (Xie, Shen, and Wang 2021). In the period from 2013 to 2016, reviewer acceptance and completion rates reached an inflection point, dipping below 50% across disciplines – a sign of possible 'reviewer fatigue' (Publons 2018, 28). Not long after, the COVID-19 pandemic unleashed an explosion of journal manuscript submissions across all fields of research (Else 2020; Squazzoni et al. 2021) and dismembered reviewer capacity as family caretaking responsibilities intensified (Squazzoni et al. 2021), triggering a full-blown 'peer-review crisis' (Künzli et al. 2022).

Meanwhile, the majority of surveyed scientists have come to believe that science is in the grip of a replication crisis (Baker 2016). Replication studies – which evaluate a previously reported scientific finding by analysing new data (Committee on Reproducibility and Replicability in Science et al. 2019, 46) – have found low rates of replication success for studies in biomedicine (Begley and Ellis 2012; Ioannidis 2005; Prinz, Schlange, and Asadullah 2011), psychology (R. Klein et al. 2014; Open Science Collaboration 2015a), and economics (Camerer et al. 2016; Chang and Li 2017). A survey of scientists

**CONTACT** Carole J. Lee c3@uw.edu Department of Philosophy, University of Washington, Box 353350, Seattle, WA 98195, USA

found that 70% had tried and failed to replicate another researcher's experimental results (Baker 2016). In an effort to improve the replicability of published research studies, journals and grant agencies have worked to incentivise transparency in research reporting, including the sharing of open science elements such as data, code, and (when appropriate) preregistered analysis plans (Begg et al. 1996; Committee on Reproducibility and Replicability in Science et al 2019; National Academies of Sciences, Engineering 2020; Nosek et al. 2015). Open science elements have been elevated into credibility signals of a sort (Jamieson et al. 2019; Lee 2022), with some evidence of increased data sharing in the social and biomedical sciences (Ferguson et al. 2023; Wallach, Boyack, and Ioannidis 2018).

Now, large-scale accessibility and improvements to generative AI empowers unscrupulous scientists to fabricate manuscript content – narrative, images, data, and code – with unparalleled speed and ease (Conroy 2023; Economist 2024). Meanwhile, predatory journals touting apocryphal peer review claims remain poised to publish such 'findings' (Moher et al. 2017). The deluge of fake papers obstructs efforts to evaluate scientific claims via systematic review, which synthesises evidence across multiple studies to provide the largest, richest inferential base (Else 2024). Fabricated studies heighten concerns about study replicability and make downstream expert evaluation of study quality even more time intensive (Else 2024).

In the face of these tangled crises in peer review capacity, study replicability, and fabricated science, the motivation for scalable tools for evaluating scientific claims has never been more intense. The surging industry of effort in this space includes: the Black Spatula project investigating the extent to which large language models (LLMs) can identify numerical, methodological, logical, citation, and other errors (The Black Spatula Project n.d.; Liu and Shah 2023; Lou et al. 2025; Zhang and Abernethy 2024); CORE-Agent, a task-specific Auto-GPT tool for evaluating computational reproducibility given accompanying content such as study data, code, and README files (Siegel et al. 2024); a machine learning model using features like word frequencies, readability scores, and formatting measures to predict peer review scores (Checco et al. 2021); and a GPT-4o-based agent which generates descriptions of conference paper strengths and weaknesses, numerical peer review scores, and preliminary outcome decisions (accept or reject) (Lu et al. 2024). Some journals recommend that authors use automated tools as checks before submission (Association for Psychological Science 2024). Others have incorporated them in the review process to increase efficiency and 'objectivity' (Frontiers Communications 2020).

Given the sociotechnical nature of the crises facing the scientific enterprise, it is imperative to anticipate how the interests, needs, and incentives of individual scientists, academic institutions, government agencies, academic societies, social and traditional media, and publishers, preprint platforms, and other organisations offering research, communication, and evaluation-related services, may shift how automated tools are implemented and interpreted. Not only may downstream use depart from their creators' intended function (Law 2012; Selbst et al. 2019; A. Wang et al. 2024). They may be repurposed in problematic ways.

In the past, scientific communities – when faced with increased pressure to evaluate and sort research and their authors – have decontextualised, repurposed, and aggregated newly crafted credibility markers (Lee 2022) through fallacious processes of inference by false ascent. In *inference by false ascent* (Cartwright et al. 2023, 48; Reiss 2016), the original, concrete purpose for which a measure is originally designed gets redescribed at a more abstract level, but in epistemically problematic ways: for example, although the journal impact factor was first introduced for the purpose of helping librarians decide which journals to purchase (Gross and Gross 1927), some embraced this metric as a more abstract measure of 'scientific impact' (Garfield 2006). Because the more abstract purpose covers a range of other more concrete purposes, the original measure may be deployed for those other concrete purposes, even when doing so is epistemically inapt: for example, the journal impact factor, once elevated into an abstract measure of 'scientific impact', was repurposed as a controversial measure of article and author impact (Hicks et al. 2015; Larivière et al. 2016) and likely informs determinations of article relevance in research search engines like Google Scholar (Google Scholar n.d.). Even open science elements shared for the purpose of facilitating the transparent evaluation of a manuscript's claims (Begg et al. 1996; Nosek et al. 2015) can be easily decontextualised and repurposed (Leonelli 2023) to score manuscripts for their computational reproducibility (Siegel et al. 2024) and, when combined with data citation metrics, to quantify author data-impact via 'S-index' metrics (Challenge.gov n.d.).

This paper aims to achieve two major goals. The first is to warn that these problems of inference by false ascent readily recur when users, designers, and stakeholders engage with AI tools for evaluating science. As a broad class, AI science evaluation tools assess scientific objects (e.g. study results, data, analytical choices, code, manuscript text, authors) for their credibility (e.g. replicability, statistical validity, rigour, computational reproducibility, impact). AI science evaluation tools are vulnerable to inference by false ascent due to disagreement about how to conceptualise and measure normative aspects of science, the portability of these tools and their outputs across purposes, and technical demands for large data sets whose features may not prioritise epistemic fit.

The second major goal of this paper is to identify epistemic resources – and a newly articulated scientific norm – that can counter inference by false ascent via AI science evaluation tools. I will explain how a pragmatist epistemology that indexes reliability assessments to *specific purposes* (Cartwright et al. 2023, 48; Reiss 2016) can surface and diagnose these epistemic pathologies. However, a pragmatist epistemology needs additional theoretical resources to adjudicate debate about the purposes and the conditions under which AI science evaluation tools are considered reliable for purpose. To address these problems, I will identify how social epistemic resources from procedural accounts of objectivity (Longino 1990, 2002) can help – and how this framework can ground a broader vision for the democratisation of the design, interpretation, use, and evaluation of these tools in scientific communities. To close, I articulate the scientific norm of *Critically Engaged Pragmatism* to enjoin scientific communities to vigorously scrutinise the purposes and purpose-specific reliability of AI science evaluation tools. To foster this norm, creators of AI tools should transparently and fully report details about how they are designed, trained, and benchmarked to facilitate these assessments; and, standards for transparent reporting should be updated as new forms of error, bias, and gamesmanship are discovered.

To illustrate these points, I will present a group of machine learning models used to predict study replicability (Yang, Youyou, and Uzzi 2020) that have gained visibility and notoriety (Crockett et al. 2023). These models depend on a method for measuring successful replication appropriate for some uses but not others; the edicts of the models are, nevertheless, interpreted as involving a more abstract quality ('replicability') that covers the broader range of uses; then, the model's imagined usefulness is expanded to cover cases for which the original measure is not apt. Focusing on supervised predictive optimisation machine learning models such as these is useful for expository reasons. First, because supervised machine learning models require the careful curation of data involving labelled outcomes of interest, we have access to an articulated justification for the chosen measures and training data. Second, these models' focus on predictive accuracy helps to illustrate why externalist epistemological frameworks are inadequate for diagnosing and addressing inference by false ascent.[1] However, I take the theoretical lessons of this paper's analysis to generalise beyond predictive optimisation machine learning tools.

## 2. Machine Learning Models for Evaluating Science

Before introducing my case study, I must first provide some background on machine learning models for evaluating science.

There are two general classes of supervised machine learning models. *Automating judgment* machine learning models are trained on *human judgments* about the outcome of interest given certain criteria. For example, an automated essay grading tool training set could include teacher scores for essays given specific features (Barocas, Hardt, and Narayanan 2023, 30). By inferring pre-existing decision-making patterns, such an approach can lead to outcomes that are less inconsistent than human judgements (Barocas, Hardt, and Narayanan 2023, 27–30). For example, a machine learning model trained on grant application abstracts and the disciplinary categories to which they had been assigned by the applicant (or to which they had been reassigned by a panel chair) was found to be more consistent in assigning disciplinary categories to abstracts than STEM PhD students and postdocs were (Goh et al. 2020). Broadly speaking, improved consistency in judgement is usually interpreted as a procedural improvement as a matter of general psychometric principle (Rust and Golombok 2014). In the context of scientific evaluation, it has also been interpreted as an improvement in human (Lee 2012; Sattler et al. 2015) and automated (Lu et al. 2024) peer review.

In contrast, *predictive optimization* machine learning models are trained on data about selected criteria and the outcome of interest and develop their own rule for predicting the outcome of interest (Barocas, Hardt, and Narayanan 2023, 28–31). For example, a tool for determining 'creditworthiness' in the context of lending could be trained on past loan default data and credit scores to predict future loan repayment (A. Wang et al. 2024). By inferring patterns between criteria and outcomes in a way that maximises predictive accuracy in ground truth data, the ambition is for predictive optimisation machine learning models to become *more reliable* than human experts (Barocas, Hardt, and Narayanan 2023, 31). For example, machine learning models for predicting the replicability of social science studies were found to have accuracy levels ranging from 0.65 to 0.78 (Yang, Youyou, and Uzzi 2020), which compares favourably to more resource-intensive, human evaluation processes such as prediction markets and surveys (Dreber et al. 2015). Broadly speaking, improved predictive accuracy is usually considered a procedural improvement as a matter of psychometric principle (Rust and Golombok 2014). In the context of scientific evaluation, predictive accuracy has also been considered an important basis for evaluating peer review's epistemic standing – for example, its ability (or not) to predict post-publication citations (Card et al. 2020; J. Wang, Veugelers, and Stephan 2017).

Conceptually, the distinction between these two types of machine learning is not always hard and fast. This is because, in prediction optimisation models, the 'objective target of interest' (Barocas, Hardt, and Narayanan 2023, 30) can involve labels that have been assigned at some point by human judgement, which can be subjective and biased (A. Wang et al. 2024). Vestiges of such subjectivity pierce the normative air shrouding appellations to 'ground truth'. Still, for the purposes of this paper's analysis, the distinction between automating judgement and predictive optimisation machine learning models remains useful insofar as it foregrounds different primary epistemic purposes: I take the former to aim primarily for *consistency* in judgement, while I take the latter to aim primarily for *predictive accuracy*.

In the machine learning literature, it is standard practice to question how well the choice of a proxy for an outcome matches the more general construct with which it is expected to be 'synonymous' (Barocas, Hardt, and Narayanan 2023, 32). For users and stakeholders using predictive optimisation machine learning models to evaluate science, this epistemic vulnerability is the jetty from which inference by false ascent embarks.

## 3. Inference by False Ascent When Predicting 'Replicability'

The conditions for inference by false ascent are rife when it comes to predictive optimisation machine learning models for evaluating science for (at least) three reasons. First, the ground truths upon which predictive optimisation machine learning models are trained involve outcomes labelled as having some epistemically desirable quality ('excellence', 'rigor', 'replicability', etc.) or not. These normative qualities and their measurement are – in principle and often in fact – open to debate and to dispute over proxy-construct fit. This is a broader problem for machine learning models that seek to evaluate normative outcomes such as fairness (Selbst et al. 2019) and online toxicity (Gordon et al. 2022).[2] Second, machine learning models and their outputs are highly portable within and across contexts, which makes it easy for users and stakeholders to misapply them in hopes of addressing real and urgent needs in the broader sociotechnical context (Law 2012; Selbst et al. 2019; A. Wang et al. 2024). Third, machine learning models require troves of data, encouraging designers to choose measures that optimise for data set size rather than epistemic fit.

To illustrate these points, consider a set of predictive optimisation models designed to assess study 'replicability' that were published as an 'exploratory investigation' in the *Proceedings of the National Academy of Sciences* (Yang, Youyou, and Uzzi 2020, 10762). These models were trained on ground truth data, obtained by the Open Science Collaboration, about 96 studies that had passed or failed replication tests in psychology (Open Science Collaboration 2015a). The models were differentiated by which manuscript features they considered: the first included narrative text only (so excluded quantitative content), the second included reported statistics, and the third included both narrative and statistical content. These models were tested for their predictive accuracy using ground truth data about replication outcomes for 317 out-of-sample replication studies in psychology and economics (Yang, Youyou, and Uzzi 2020, 3 of supplement; Registered Replication Reports 2025; R. A. Klein et al. 2014, 2018; Ebersole et al. 2016; Nosek and Lakens 2013; Aarts et al. 2016; Openingscience.Org 2025; Höffler 2017; Camerer et al. 2016). The accuracy levels of these models ranged from 0.65 to 0.78. In comparison, after pooling data from five projects (Camerer et al. 2018; Dreber et al. 2015; Ebersole et al. 2020; Forsell et al. 2019), the accuracy levels for prediction markets and surveys were estimated to be lower at 0.52 and 0.48 respectively (Nosek et al. 2022, figure 2).

Yang et als machine learning models chose to operationalise successful replication in terms of 'the replication team's summary judgement – their subjective expert opinion – of whether the study replicated or did not replicate' (Yang, Youyou, and Uzzi 2020, supplement page 2). In a follow-up paper, they justified this choice to 'remain grounded in the human expertise' about replication outcomes (Youyou, Yang, and Uzzi 2023, 7). However, looking more closely at the data, it turns out that replication team members overwhelmingly determined that a study successfully replicated in cases where a replication had statistically significant results in the same direction as the original study (Open Science Collaboration 2015b), which suggests that the latter measure captures the epistemic basis for replication team member judgements.

Epidemiologists have contested this latter way of measuring replication success, arguing that it would be better to combine data from replication and original studies to see whether the accumulated evidence is statistically significant in the direction of the original effect. This 'total evidence' approach to measuring replication success better addresses questions about how much confidence we can have in a scientific claim, which is typically used to ground evidence-based policy and intervention recommendations (Gilbert et al. 2016; Goodman 2019; Goodman, Fanelli, and Ioannidis 2016, 3). It is important to emphasise how different these approaches to conceptualising and operationalising successful 'replication' are. A replication that fails to find a statistically significant result in the same direction as an original effect can nevertheless provide evidence that supports the original claim: for example, if a follow-up study does not replicate an original study's statistically significant result because of its smaller sample size, but the replication study's effect size mirrors that found in the original study, then the 'failed' replication can be interpreted as corroborating the original effect (Goodman, Fanelli, and Ioannidis 2016). Additionally, a replication that successfully mirrors an original study's null result can nevertheless provide evidence that, from an accumulated evidence perspective, undermines the original study's claim: multiple null result replications of an originally discovered null result, when pooled, can reveal a statistically significant effect, as was found in a meta-analysis of tamoxifen's effect on breast cancer survival rates (Goodman, Fanelli, and Ioannidis 2016).

In recognition that there is 'no single standard for evaluating replication success' (Open Science Collaboration 2015a, 943) – but instead a 'variety of options' none of which 'is definitive' (Nosek et al. 2022) – the Open Science Collaboration's mass replication study (from which Yang et al drew their training data) reported results across multiple ways of operationalising replication, including the total evidence approach:

- 36% of the replication studies had statistically significant results in the same direction as the original study;
- 39% of the replication studies were subjectively rated by the replication team as having successfully replicated the original study results;
- 47% of the original effect sizes were in the 95% confidence interval of the replication effect size; and,
- 68% of the studies had statistically significant effects in the direction of the original study when the replication and original data were combined.

We would not expect machine learning models trained with data labelled using Yang et al's preferred measure for replicability (which had a 39% success rate in the training data) to be reliable for predicting replicability according to a total evidence view (which had a much higher 68% success rate in the training data).

Despite normative disagreement about what replicability means and how best to measure it – as well as empirically demonstrated differences between these measures – Yang et al elevate their contested measure into a broader, more abstract quality of 'replicability': they state that their machine learning model 'predicts replicability' (Yang, Youyou, and Uzzi 2020, 10762). This more abstract concept of 'replicability' covers a broad range of concrete use cases, including evaluating how much confidence we can have in a scientific claim. However, for that use case, the measure that replication teams appeared to implement is not epistemically appropriate. Nevertheless, Yang et al advocate for this expanded application for their machine learning model: they characterise their tool as something that can improve '*confidence in findings*' (Yang, Youyou, and Uzzi 2020, 10767, italics mine), thus completing the inference by false ascent.

Pressing needs in the broader sociotechnical context encourage this kind of inapt repurposing. The proliferation of fabricated science and predatory journals have heightened concerns about how much confidence we can have in study results reported in preprints and published articles, which raises downstream concerns about the strength of the available evidence-base for prospective policy and intervention recommendations. To address this in the domain of national security, the Defense Advanced Research Projects Agency (DARPA) solicited proposals that would develop 'automated tools to assign Confidence Scores' to studies to help federal agencies adopt 'appropriate levels of confidence in a claim' (Chawla 2018) for the sake of 'modeling, planning for, and operating in, the Human Domain' (Defense Advanced Research Projects Agency 2018, 7). Reporting on the DARPA funding opportunity, a science news journalist described the prospective projects as future 'BS detector[s]' (Rogers 2017) – again, a reflection of broader social and scientific pressure to find ways to evaluate confidence in scientific claims.

Finally, technical demands that motivate designers to choose measures that optimise for data set size rather than epistemic fit can lead to inference by false ascent. In their supplement, Yang et al observe that replication team judgement about successful or unsuccessful replication was 'a common metric reported in each replication study' (Yang, Youyou, and Uzzi 2020, supplement page 2). By choosing this metric for measuring 'replicability', they were able to expand their combined training and test data sets to include 413 studies. Yang et al recognise that the size of even this combined data set was 'meager by most machine learning study standards' (Yang, Youyou, and Uzzi 2020, 10763): Crockett et al estimate it to be 'orders of magnitude smaller than those used to train machine learning models for far simpler tasks' (Crockett et al. 2023, 1) – small enough to possibly inflate estimates of model accuracy 'just as underpowered samples inflate false-positive findings' (Crockett et al. 2023, 1). Yang et al's choice, which optimised for data set size, laid the path for inference by false ascent by implicitly adopting a measure not reliable for a purpose they aimed to serve.

In the face of normative disagreement about the purposes to which automated tools are put, their portability across purposes, and technical demands that can motivate prioritising data set size over epistemic fit, we need theoretical resources and norms that can surface, diagnose, and counteract the epistemic pathologies wrought by inference by false ascent.

## 4. A Social, Pragmatist Epistemology for AI Science Evaluation Tools

What epistemic resources are implicitly available to a framework that prioritises predictive accuracy? And why are these inadequate for identifying and addressing inference by false ascent?

As mentioned earlier, an important ambition for predictive optimisation machine learning models is that they be *more reliable* than human experts (Barocas, Hardt, and Narayanan 2023, 31): e.g. Yang et al's machine learning models had higher predictive accuracy than human judgements aggregated by means of prediction markets and surveys (Yang, Youyou, and Uzzi 2020; Nosek et al. 2022, figure 2).

What does it mean for predictive optimisation machine learning models to be considered 'reliable'? For epistemologists, the most natural way to interpret reliability may be externalist. Such a view deems a human belief justified when it is produced by a cognitive process that produces a high ratio of true beliefs (Goldman 1986). By extension, an externalist interpretation of reliability for a machine learning model attributes knowledge in cases where it arrives at a true representation (e.g. an output measure, score, claim) by means of a process producing 'a high proportion of accurate representations' (Humphreys 2004; 2020, 24). On the surface, this externalist interpretation comports with the practice of evaluating predictive optimisation machine learning models by how well they predict ground truths outside of their training set.

In contrast to an externalist epistemology, a pragmatist epistemology evaluates scientific tools – be they measures, methodologies, instruments, analyses, models, simulations, procedures, experiments, expert assessments, or automated tools – as reliable *for achieving a specific purpose* (Cartwright et al. 2023, 11; Reiss 2016, 3). The better measure or model is the one that more effectively serves the purpose at hand (Reiss 2016, 29–30). For example, there are multiple indexes that could be used to measure 'the' Consumer Price Index (CPI). The cost-of-living index (COLI) framework allows consumers to substitute more expensive types of goods for less expensive ones that provide the same level of utility. In contrast, the Laspeyres price index tracks the cost of a fixed basket of goods over time. These measures are better suited for different purposes: the COLI framework answers questions about how much a consumer's income would need to change from one period to another in order to achieve the same level of *utility* while the Laspeyres price index answers questions about how much a consumer's income would need to change to purchase a *fixed basket of goods* (Reiss 2016, 24–27). This makes COLI more appropriate for updating Social Security benefits while the Laspeyres approach may be more appropriate for estimating basic living costs (Reiss 2016, 31). What is epistemically important is not just that the COLI framework provides a lower CPI estimate than the Laspeyres price index, but that the choice of index should be driven by purpose-specific reliability – not by judgements about which measure best matches 'the' CPI as an abstract construct (Reiss 2016, 24–27).

From a pragmatist perspective, a machine learning model and its choice of measure may be reliable for a particular purpose but not for a different one – a distinction critical for surfacing and diagnosing inference by false ascent. Recall that in Yang et al's training data, the replication teams' presumed measure for replication (statistically significant results in the same direction as the original study) is reliable for achieving one purpose (to evaluate whether replication results achieve the 'same result' as the original study). However, Yang et al elevate this measure into a more abstract quality (e.g. 'replicability'), then propose using it for a different purpose for which it is not reliable (e.g. evaluating confidence in a scientific claim). This example demonstrates how attending to the purpose-specific reliability of automated tools is critical in the face of disagreement about the purposes to which they should be put, their portability across purposes, and potential mismatch between their purpose and training. In contrast, an externalist approach, which is unattuned to purpose, overlooks the epistemic pathologies of inference by false ascent resulting from these challenges.

However, a pragmatist account remains incomplete if it does not provide theoretical resources for adjudicating debates about the conditions under which measures, models, or tools are deemed reliable for purpose or not.[3] This is critical in the context of debates about inference by false ascent as there is no fallacy when a measure, model, or tool used for an original purpose *is* reliable for other purposes. Pragmatism provides some help: for pragmatists, ascertaining the purpose-specific reliability of any scientific tool must be 'vouchsafed' by methods considered reliable for that purpose (Cartwright et al. 2023, 30). Such judgements invoke 'background knowledge' (Reiss 2016, 6) – or, more colourfully, an entire 'tangle of science' involving a 'vast network of other scientific work that defines the concepts used in them, develops measures for them, assesses the validity of these measures, tests the claims implied once the concepts are well defined, and much more' (Cartwright et al. 2023, 30).

However, appealing to a shared body of knowledge may not be sufficient for mediating disagreement about the purpose-specific reliability of a measure, model, or tool. For example, imagine that replication teams argue that measuring replications in one way (statistically significant results in the same direction as the original study) for some original purpose (to evaluate whether replication results mirror those from the original study) *is reliable* for addressing some second purpose (evaluating confidence in a scientific claim). They may believe (with evidence) that journals prefer to publish statistically significant results, which induces authors to selectively report or engage in questionable research practices that inflate the rate of false positives in the published literature. As such, they take data from replication studies to be less biased than data from original studies, which is why they prefer not to pool replication and original study data together when evaluating confidence in an original study's claim. In this dialectic, there is a lot of agreement – a shared evidence base (about how journals and authors behave as well as specific results from replication and original studies) as well as shared standards of inference supporting that evidence base. Yet, different camps draw different inferences about which measure is more reliable for the purpose at hand.

This epistemic situation – in which scientists draw different inferences despite shared evidence and standards of inference – occurs across a broad range of scientific activities. To address this kind of underdetermination of inference by evidence, procedural accounts of objectivity appeal to social epistemic resources that aim to foster critical discursive interactions that push back against idiosyncratic judgements (Longino 1990, 2002). In this approach, Helen Longino advocates for four norms (Longino 1990, 2002), which I present here in broad strokes. First, scientific communities should have public venues that allow researchers to share their work as well as their critiques of the evidence, methods, and inferences employed by others. Second, criticism should receive uptake: in response to criticism, scientists should develop new data, methods, or arguments to support their views or modify their beliefs. Third, scientists should hold shared standards by which 'points of agreement, points of disagreement, and [grounds for] what would count as resolving the [latter] and destabilizing the former' can be identified (Longino 2002, 130–31).[4] Fourth, scientists should foster critical engagement with a diversity of perspectives to ensure that scientific inferences and standards are tested against a broad range of social, economic, and political perspectives. Adherence to these norms cannot guarantee that disputes about purpose-specific reliability will always be resolved. However, it can foster the kind of critical discursive practices that impede idiosyncrasy in judgement and encourage progress through the development of new data, methods, and arguments. As such, a pragmatist epistemology for automated tools would be improved by embracing social epistemic resources provided by procedural accounts of objectivity.

Moreover, procedural accounts of objectivity can provide resources for adjudicating and mediating debates about the *purposes* to which machine learning models and their measures should be put. When presenting his pragmatist view, Julian Reiss suggests that the purposes for measures used to inform evidence-based policy (e.g. whether a CPI measure should address how much a consumer's income would need to increase to achieve the same level of utility versus purchase a fixed basket of goods) should be determined by the body politic (Reiss 2016). In contrast, in her account of procedural objectivity, Longino suggests that critical discursive engagement *within scientific communities* can include disputes about 'the validity of cognitive standards and of *cognitive goals*' (Longino 2002, 164, italics mine). Reiss's and Longino's views have different implications for what it would mean to democratise AI science evaluation tools. In Reiss's view, democratisation would require identifying the purposes to which machine learning models and their measures should be put according to the broader community within which scientific communities are embedded. In Longino's view, democratisation would require fostering debate *within* scientific communities about the purposes, design, interpretation, use, and evaluation of automated tools: scientists – in their roles as designers, users, and stakeholders – would be obliged to take up criticism related to epistemic pathologies wrought by model decontextualization, repurposing, and implementation.

Overall, a pragmatist epistemology conjoined with a procedural account of objectivity can surface and diagnose inference by false ascent while democratising debate and uptake about the purposes and purpose-specific reliability of AI science evaluation tools within scientific communities. Moreover, such a framework provides social epistemic resources for countenancing criticism about biases, inaccuracies, conservativism, and homogenisation in AI science evaluation tool outputs (Birhane et al. 2022; Crockett et al. 2023; Messeri and Crockett 2024; Resnick and Hosseini 2025; Resnick, Hosseini, and Hauswald 2025; Whaanga 2020), questions about whether 'reliability' should be measured by giving 'equal weight to each instance' or in a way that does not deprioritize 'those underrepresented in the data or the world' (Birhane et al. 2022, 179), and disparate harms AI tool designs may have on the research and epistemic standing of scientists belonging to historically marginalised groups (Birhane et al. 2022; Messeri and Crockett 2024). The epistemic and financial priorities of the most powerful and well-resourced, including AI tool creators, should not be entrenched (Birhane et al. 2022; Lazar and Nelson 2023; Selbst et al. 2019) in ways that thwart norms fostering critical discursive engagement, including perspectival diversity and critical uptake.[5]

## 5. Conclusion: Critically Engaged Pragmatism

In the mid-twentieth century, journals like *Science* and the *American Journal of Medicine* adopted peer review to manage editorial workload (Baldwin 2018). Since then, the number of manuscripts has grown exponentially, threating to overwhelm reviewer capacity at a time when concerns about study replication and AI-fabricated science have intensified. The need for scalable tools for evaluating the credibility of science and scientists – and the need for epistemological frameworks and norms for assessing AI science evaluation tools – has never been more acute.

In this paper, I caution that, when faced with increasing pressure to evaluate scientific research and its authors, science has a history of decontextualising and repurposing credibility markers in epistemically fallacious ways – and that AI science evaluation tools are vulnerable to the same kinds of inference by false ascent due to disagreement about how to conceptualise and measure normative aspects of science, their portability across purposes, and technical demands that prioritise data set size over epistemic fit. To counter this threat, I argue for a social, pragmatist epistemology that fosters and democratises critical engagement within scientific communities about the purposes to which AI science evaluation tools should be put and their purpose-specific reliability.

Insofar as norms of science serve as 'bulwarks' against contemporary threats to 'the extension of certified knowledge' (Lee 2022, 1003) – and insofar as this social, pragmatist epistemology helps to diagnose and address inference by false ascent at the hands of AI science evaluation tools during a time of structural strain in the scientific community – I propose the norm of *Critically Engaged Pragmatism* to enjoin scientific communities to scrutinise the purposes and purpose-specific reliability of AI science evaluation tools. To foster Critically Engaged Pragmatism, it is imperative that the creators of AI tools transparently and fully report details about how they are designed, trained, and benchmarked to facilitate assessments of their purpose-specific reliability, liability to different kinds of error, and bias.[6]

What count as best practices for the transparent reporting for AI science evaluation tools should be developed to facilitate independent evaluation and critique. As with standards for transparent reporting in the sciences – e.g. the Consolidated Standards of Reporting Trials (CONSORT) Guidelines, the Transparency and Openness Promotion (TOP) Guidelines – journals and conference proceedings should choose to endorse standards they identify as epistemically apt for their venue.[7] However, unlike the CONSORT and TOP Guidelines, standards for the transparent reporting of AI science evaluation tools should be updated at faster intervals, perhaps in a living document, as new forms of bias, error, and gamesmanship are uncovered.

Although the full implementation of Critically Engaged Pragmatism is currently constrained by structural forces, the social, pragmatist epistemology motivating its adoption illuminates directions towards which individuals, institutions, and communities should strive. For example, although science evaluation tools based on commercial technologies are not fully accessible or responsive to scientific critique (Messeri and Crockett 2024), a growing body of open-weight (HuggingFace 2026) and trained LLMs (Asai et al. 2026) provides an opportunity for the scientific community to design and evaluate LLM-based agents for which they have increased transparency and control; and, although technical expertise may be difficult to access due to disciplinary and academia-industry divides (Seger et al. 2023), Critically Engaged Pragmatism spotlights the need for scientific communities to foster relevant skills and cross-disciplinary engagement to promote informed cross-field critique, uptake, and design. Such efforts are critical because AI science evaluation tools are not the objective arbiters of scientific credibility. Rather, they are the object of the kind of critical discursive practices that ultimately ground the credibility of scientific communities.

## Notes

1. Note that an implicit externalist epistemological framework can be found in AI research outside of the predictive optimisation machine learning model literature: for example, large language models used to synthesise the scientific literature have been compared for their correctness (Asai et al. 2026).
2. To better accommodate complexity and heterogeneity, scholars belonging to historically marginalised groups have called for machine learning models and labels 'that allow for non-binaries, plural truths, contextual truths, and many ways of being' (Birhane et al. 2022, 179; Constanza-Chock 2018; Hamidi, Scheurman, and Branham 2018; Lewis et al. 2020).
3. This problem is analogous to the canonical critique against externalism that, without theoretical resources for identifying the conditions under which some process is deemed reliable, externalism is incomplete (Longino 2002, 163; Pollock 1984).
4. This focus on shared standards of reasoning distinguishes procedural accounts of objectivity from research exploring technical means for capturing and/or aggregating dissensus and pluralism among a diversity of perspectives about the proper application of normative concepts/labels but without reference to their underlying reasons (e.g., Cabitza, Campagner, and Basile 2023; Gordon et al. 2022; Sorensen et al. 2024).
5. In addition to epistemic obligations, scientists may also have ethical responsibilities to work with impacted 'communities, populations, and other stakeholders' to address bias-related concerns (Resnick and Hosseini 2025).
6. Such details are especially vital when AI tools travel across disciplinary boundaries, as 'the largely invisible "researcher degrees of freedom"' embedded in their design may reflect epistemic assumptions and methodologies inappropriate for other fields and purposes (Messeri and Crockett 2024, 54).
7. Analogously, efforts to articulate best practices in transparent reporting have started to emerge in the context of AI-*generated* research: for example, Ziming Luo, Atoosa Kasirzadeh, and Nihar Shah 'recommend that journals, conferences, and other research evaluation bodies require the submission of complete log traces and code alongside any AI scientist–generated manuscript' and 'that creators of automated AI research systems release these artifacts' (2025, 18) to allow improved detection of pitfalls such as post-hoc selection bias, data leakage, and dataset fabrication.

## Acknowledgments

An early version of this paper was presented at the 'Openness and Inequity in Research' Symposium organized by Sabina Leonelli for the 2024 Philosophy of Science Association Meeting. Many thanks to the audience, Molly Crockett,

Mohammad Hosseini, Shahan Memon, David Resnik, Nihar Shah, Daniel Scott Smith, Jevin West, and the anonymous reviewers for comments that sharpened earlier versions of this paper. All views (and epistemic foibles) are those of the author.

## Disclosure Statement

No potential conflict of interest was reported by the author(s). CJL served on the Coordinating Committee and the Advisory Board of the Transparency and Openness Promotion (TOP) Guidelines.

## Notes on contributor

***Carole J. Lee*** studies how scientific institutions award and set standards for markers of credibility such as successful peer review outcomes. Her research has been published across fields and supported by the National Institutes of Health, the National Science Foundation, and a Career Enhancement Fellowship funded by the Mellon Foundation and administered by the Institute for Citizens & Scholars. She received the Philosophy of Science Association's Prize in Philosophy of Science & Race and, with Elena Erosheva, received first prize for most creative submission to the National Institutes of Health's Peer Review Challenge.

## ORCID

Carole J. Lee http://orcid.org/0000-0002-5323-9205